# Combination of Thermal Subsystems Modeled by Rapid Circuit Transformation

Y.C. Gerstenmaier*, W. Kiffe*, and G. Wachutka**

*Siemens AG, Corporate Technology, 81730 Muenchen, Germany, e-mail: yge@tep.ei.tum.de
**Institute for Physics of Electrotechnology, Munich University of Technology, Germany

*Abstract-* **This paper will deal with the modeling-problem of combining thermal subsystems (e.g. a semiconductor module or package with a cooling radiator) making use of reduced models. The subsystem models consist of a set of Foster-type thermal equivalent circuits, which are only behavioral models. A fast algorithm is presented for transforming the Foster-type circuits in Cauer-circuits which have physical behavior and therefore allow for the construction of the thermal model of the complete system. Then the set of Cauer-circuits for the complete system is transformed back into Foster-circuits to give a simple mathematical representation and applicability. The transformation algorithms are derived in concise form by use of recursive relations. The method is exemplified by modeling and measurements on a single chip IGBT package mounted on a closed water cooled radiator. The thermal impedance of the complete system is constructed from the impedances of the subsystems, IGBT-package and radiator, and also the impedance of the package can be inferred from the measured impedance of the complete system.**

## I. INTRODUCTION

For the calculation of hot spot temperatures or temperature fields in electronic systems with rapidly varying chip heat source strength usually reduced models are used. Numerous compact static [1-5] and transient [6-10] thermal models have been established for a rapid calculation of temperatures. The notion of "compact" thermal model usually implies boundary condition independence (BCI) [1, 2], i.e. the model is valid for all (or nearly all) reasonable temperatures, heat flows and also heat transfer coefficients applied to the thermal contact areas. An advantage of the model presented in [11, 12] is its ease of parameter determination by simple linear least square fit to measured or simulated heating curves (thermal impedances). The model was extended in [12] to include the effects of varying surface or ambient temperature and varying heat flows at the thermal contact areas. However, it is not possible to use arbitrary heat transfer coefficients α as external bound.c. parameters independently of the model parameters. Thus new model parameters have to be determined, when α changes. On the other hand the compact models of [1, 2, 4, 5] deal with small packages with small thermal contact areas in comparison to the multi chip module of Fig. 1, which is mounted on a cooling radiator. Such modules always have a large temperature variation along the bottom side of the module base plate depending on the different heating cases. Thus the module can hardly be approximated thermally by a compact model which is independent of α.

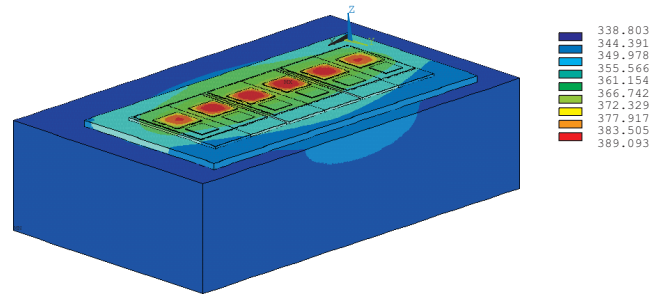

Fig. 1: FEM-simulation of steady state temperature (°K) distribution for 1700V IGBT and diode module mounted on cooling radiator.

Most of the reduced models, also [11, 12], apply to systems with negligible nonlinearities, e.g. the material parameters are supposed to be temperature independent just as is α. Methods for reducing nonlinear systems are presented e.g. in [13, 14].

This paper will deal with the modeling-problem of combining thermal subsystems, e.g. combining a power semiconductor module with a cooling radiator, making use of the reduced models of [11, 12]. Heat transfer coefficients α in this context can also be modeled as thermal subsystem (layer of thermal resistance).

## II. THERMAL MODEL

The model in [11, 12] characterizes the thermal set-up by a set of $M \leq 20$ effective time constants $t_i$ which are logarithmically distributed, typically between $\text{Min}(t_i) = 10^{-4}$ s and $\text{Max}(t_i) = 1000$ s (depending on system and heat source size), and besides this are chosen freely. The model equation for the temperature field $T(x, t)$ reads:

$$T(x,t) = \sum_{i=1}^{M} \sum_{l=1}^{L+C+J} \mathfrak{M}_{i\,l}(x) \cdot \int_0^t s_l(\tau)\, e^{(\tau-t)/t_i}\, d\tau \qquad (1)$$

where $s_l(t)$ denotes a heat source term which can be either the dissipated power $p_l(t)$ of a chip $l$, $l = 1, .., L$ or an applied average ambient temperature $T_{a,c}(t)$ at a thermal contact area $c = 1, .., C$ or a thermal heat flux $J_k(t)$ at a thermal contact $k = 1, .., J$. The matrices $\mathfrak{M}_{i\,l}(x)$ represent the model parameters for a chosen set of locations $x$, usually the hot spots of the system in the chip centers. The $\mathfrak{M}_{i\,l}(x)$ values are obtained by linear least square fits to unit step responses of FEM-simulated or meas-ured $T(x, t)$ for individual $s_l(t)$.

Using the expression (1) for single source unit-step heating with $s_l(t) = \Theta(t)$ and $s_m(t) = 0$ for all $m \neq l$ the temperature field in case of homogeneous starting temperature $T(x,0) = 0$ takes the more customary appearance of a thermal impedance (heating curve for unit power):

$$zth_l(x,t) = \sum_{i=1}^{M} R_i(x)(1 - e^{-t/t_i}) \qquad (2)$$

with $R_i(x) = t_i \mathfrak{N}_{i\,l}(x)$. $zth_l(x,t)$ denotes the unit step response of the system for heating only e.g. chip $l$ with unit strength of power or one thermal contact area with a unit step in ambient temperature or one thermal contact with a unit heat flux step. $zth_l(x,t)$ thus is the generalized thermal impedance for fixed position $x$. It can serve to calculate $T(x,t)$ for arbitrary time evolution $s_l(t)$ by convolution of $s_l(t)$ with the time derivative $\dot{z}th_l(x,t)$ [15, 16], which can also be directly inferred from (1). Having simultaneously several sources $s_l(t) \neq 0$ the individual contributions to the temperature field are superposed according to (1).

Eq. (2) is represented by the Foster thermal circuit of Fig. 2 with $R_i C_i = t_i$, thus $C_i = t_i /R_i$. Since (1) and (2) are valid models for general 3D systems (with negligible nonlinearities) the circuit of Fig. 2 can also represent 3D systems. Superposition of several heat sources can be represented by a series connection of circuits of Fig. 2 [11] together with their individual heat sources (pseudo-current sources).

When two or more thermal systems are combined, it is not straightforward to derive the thermal model of the combined system from the known thermal models of the individual systems. A typical task in practice is to mount a semiconductor module or single chip package on a cooling radiator and to provide the thermal model for the complete system. The model for the combined system is not obtained by series connection of the Foster-circuit junction-case for the package and the Foster-circuit case-ambient for the radiator, because the Foster-circuit is only a behavioral description of the subsystems but no true physical description. This is easily recognized by the fact that heat propagation through the Foster circuit is instantaneous, i.e. the heat flow entering at the left hand (source) side leaves at the same moment with equal strength at the right hand side by current continuity and would flow into the series connected radiator circuit. In reality it needs some time for the package to warm up until heat flow into the radiator occurs.

The Cauer-circuit shown in Fig.3 can account for this delay and provides a much more physical description of the heat flow path and it can describe the same thermal impedances $zth(t)$ as the Foster-circuits in case of constant right hand side temperature. The Cauer-circuit is closely related to the concept of "structure function" [17]. The Cauer-circuits being physical models can be series connected to form Cauer ladders for the total system. When connecting the Cauer-circuits, they actually form two-ports contrary to the one-port Foster-circuits [8], Fig.9. However, the mathematical representation of the Cauer form is much more complicated than the Foster eq. (2) and it is more difficult to determine the Cauer network parameters $r_i$, $c_i$ by fit. For this reason the thermal impedances in module data sheets are provided by Foster-parameters, which are of little use for combined systems.

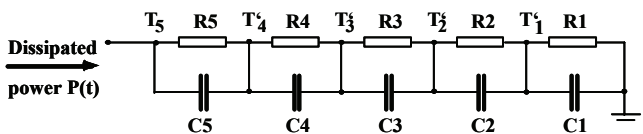

Fig. 2: Foster type thermal equivalent circuit with applied heat power as "current"-source.

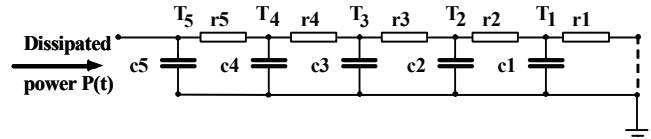

Fig. 3: Cauer type thermal equivalent circuit. When omitting the dashed line, the circuit forms a two-port for connection to another sub-system Cauer-circuit.

The method suggested in this work proceeds in three steps: First the (case-) interface between the two subsystems (e.g. module and radiator) is subdivided into several thermal contact areas, so that the temperature at each contact area is approximately homogeneous. This will be necessary for large modules as in Fig. 1. The thermal subsystems are described by Foster-models of the form (1), (2). In a second step for each thermal contact area the corresponding Foster-circuits in subsystem 1 and 2 are transformed in Cauer ladders and connected at the thermal contacts. The result is a parallel connection of Cauer-ladders between junction (module chip heat source) and ambient (bottom of radiator or cooling fluid with constant ambient temperature). The parallel Cauer ladders are combined in the Laplace-domain by straightforward algebra to form one thermal impedance, which is transformed in a third step to a Foster-circuit junction-ambient by partial-fraction-decomposition (section III). The Foster circuits thus obtained characterize the combined system thermally by simple mathematics of the type of (2).

III. TRANSFORMATION OF EQUIVALENT THERMAL CIRCUITS

The Foster-Cauer circuits are treated in works on network synthesis [18]. In the following a concise derivation for the Foster - Cauer transformation and vice versa will be presented, which leads to fast algorithms so that a large number of transformations can be performed, sufficient for the description of whole temperature fields. The Foster-circuit of Fig. 2 can be created recursively (or iteratively) by the prescription expressed in Fig.4. $Zth_n(t)$ is the one-port impedance of the circuit with $n$ thermal resistors and capacitors $R_1, .., R_n, C_1, ..., C_n$. The initial impedance $Zth_0(t)$ in Fig. 4 and 5 is zero, i.e. $Zth_0$ has zero resistance (direct connection). Generally, the effect of a one-port impedance with constant reference temperature $T_0$ on the right hand side and temperature $T_n(t)$ on the left hand side where a time dependent "current"-source $P_n(t)$ is applied, can be represented for the initial condition $T_n(t = 0) = T_0$ by the convolution integral:

$$T_n(t) - T_0 = \int_0^t P_n(\tau) Zth_n(t-\tau) d\tau \quad (3)$$

$Zth_n(t)$ as used here is the "impulse"-response (δ-response) and equal to the time-derivative of the unit-step response $zth(t)$ of (2). The impedance of a simple thermal resistor $R$ without capacitance is in this notation $Zth(t) = \delta(t) R$.

The circuit of Fig. 4 leads to the following equation system for $T_0 = 0$ and $ZthF_n(t)$ denoting the Foster-type impedances:

$$P_n(t) = P_{n-1}(t) = (T_n(t) - T_{n-1}(t))/R_n + C_n \frac{d}{dt}(T_n(t) - T_{n-1}(t))$$

$$T_n(t) = \int_0^t P_n(\tau) ZthF_n(t-\tau) d\tau, \; T_{n-1}(t) = \int_0^t P_{n-1}(\tau) ZthF_{n-1}(t-\tau) d\tau$$

This determines $ZthF_n(t)$ when $P_n(t)$ and $ZthF_{n-1}(t)$ are known. The integro-differential equation system is transformed into an algebraic system by applying the Laplace-transformation

$$\mathscr{L}\{T(t)\} = T(s) = \int_0^\infty T(t) e^{-st} dt$$

with $s = i\omega$. Under the linear operation $\mathscr{L}$ the convolution integrals are transformed in simple products $P_n(s) Zth_n(s)$ in s-space and time derivates transform to products with $s$: $\mathscr{L}\{dT(t)/dt\} = s\, T(s)$. The Laplace-transformed equation system for Fig.4 reads:

$$P_n(s) = P_{n-1}(s) = (T_n(s) - T_{n-1}(s))/R_n + C_n s (T_n(s) - T_{n-1}(s))$$

$$T_n(s) = P_n(s) ZthF_n(s), \; T_{n-1}(s) = P_{n-1}(s) ZthF_{n-1}(s)$$

From this:

$$T_n(s)/P_n(s) = ZthF_n(s) = 1/(s C_n + 1/R_n) + ZthF_{n-1}(s)$$

and the $ZthF_n(s)$ can be calculated very simply recursively to yield the well known form [19]:

$$ZthF_N(s) = \sum_{k=1}^N 1/(C_k(s-s_k)), \; s_k = -1/t_k = -1/(R_k C_k) \quad (4)$$

The Cauer-circuit of Fig. 3 is generated by the recursive prescription of Fig.5, which is expressed analytically:

$$P_{n-1}(t) = (T_n(t) - T_{n-1}(t))/r_n, \; P_n(t) = P_{n-1}(t) + c_n \frac{d}{dt} T_n(t)$$

$$T_n(t) = \int_0^t P_n(\tau) ZthC_n(t-\tau) d\tau, \; T_{n-1}(t) = \int_0^t P_{n-1}(\tau) ZthC_{n-1}(t-\tau) d\tau$$

and Laplace transformed reads:

$$P_{n-1}(s) = (T_n(s) - T_{n-1}(s))/r_n, \; P_n(s) = P_{n-1}(s) + c_n s\, T_n(s)$$

$$T_n(s) = P_n(s) ZthC_n(s), \; T_{n-1}(s) = P_{n-1}(s) ZthC_{n-1}(s)$$

From this the recursive definition of the Cauer impedances $ZthC_n(s)$ results:

$$T_n(s)/P_n(s) = ZthC_n(s) = 1/(s c_n + \frac{1}{r_n + ZthC_{n-1}(s)}) \quad (5)$$

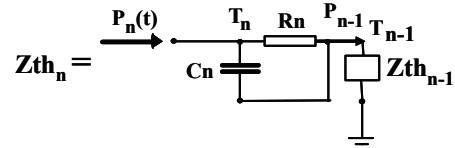

Fig. 4: Prescription for recursive generation of Foster-circuit of Fig.2

Eq. (5) leads to a continued fraction representation of $ZthC_n(s)$ which can also be written as rational function in $s$, $p_n(s)/q_n(s)$, with polynomial degree of $p_n(s)$ smaller by one than that of polynomial $q_n(s)$. An efficient algorithm for a fast calculation of the $p_n(s), q_n(s)$ can easily be derived.

The Cauer-impedance $ZthC_n(s) = p_n(s)/q_n(s)$ can be transformed into the Foster-form (4) by decomposition of the rational function in partial fractions. At first the zeroes $s_k = -1/t_k = -1/(R_k C_k)$ of $q_n(s) = 0$ have to be determined, which can be done symbolically up to $n \leq 4$ degree and numerically without limitation in $n$. The coefficients in (4) can be shown to be $1/C_k = p(s_k)/q'(s_k)$. Thus the Foster $R_k, C_k$ are obtained from the Cauer $r_k, c_k$. The symbolic calculation of the Foster $R_k, C_k$ (limited to $n \leq 4$) gives rise to extremely lengthy, unwieldy expressions that will not be reproduced in this paper, however the numerical evaluation is done very quickly and precisely.

The inverse transformation of the Foster-circuit into the Cauer-circuit makes use of the recurrence relation (5) in the form:

$$1/ZthC_n(s) = s\, c_n + 1/(r_n + ZthC_{n-1}(s)) \quad (6)$$

Equating the Foster-impedance (4) written as rational function $ZthF_n(s) = p_n(s)/q_n(s)$ for $n = N$ with the Cauer $ZthC_n(s)$, the rational function $1/ZthC_n(s) = q_n(s)/p_n(s)$ is decomposed by the standard Euklid's algorithm into a polynomial linear in $s$ and a rational function $rem_n(s)/p_n(s)$ as remainder:

$$1/ZthC_n(s) = q_n(s)/p_n(s) = s\, c'_n + k_n + rem_n(s)/p_n(s)$$

The polynomial degree($rem_n$) < degree($p_n$). Comparing this expression with (6), $c'_n$ has to be identified with $c_n$ and $k_n + rem_n(s)/p_n(s)$ with $1/(r_n + ZthC_{n-1}(s))$. Making use of the identity

$$k_n + \frac{rem_n(s)}{p_n(s)} = 1/\frac{p_n(s)}{k_n p_n(s) + rem_n(s)},$$

we have:

$$p_n(s) / (k_n p_n(s) + rem_n(s)) = r_n + ZthC_{n-1}(s).$$

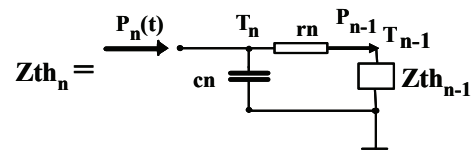

Fig. 5: Prescription for recursive generation of Cauer-circuit of Fig.3

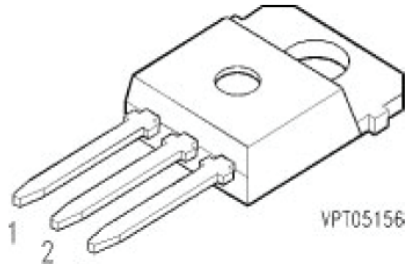

Fig. 6: TO-218AB Package for 1200V / 35A Siemens/Infineon single chip IGBT, BUP 307

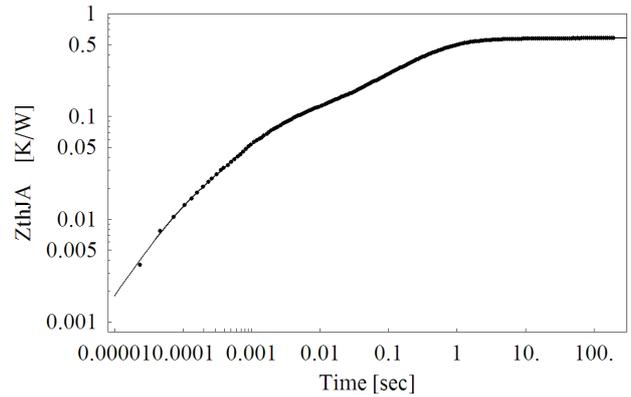

Fig. 7: Thermal impedance junction-ambient ZthJA for 1200V/35A IGBT mounted on radiator. Dissipated power 14.73W. Curve fit with model (2) to data from measuring signals averaged in small time intervals.

Using Euklid's decomposition for the rational function on the left hand side and setting $ZthC_{n-1}(s) = p_{n-1}(s) / q_{n-1}(s)$ the following relations are obtained:

$r_n = 1/k_n$ , $q_{n-1}(s) = k_n p_n(s) + rem_n(s)$, $p_{n-1}(s) = -rem_n(s)/k_n$ .

Thus the Cauer $r_n$ , $c_n$ are determined for $n = N$. The new $ZthC_{n-1}(s)$ defined by $p_{n-1}(s)/q_{n-1}(s)$ can be used in the same way as $ZthF_n(s) = p_n(s)/q_n(s)$ above, in order to determine the next Cauer $r_{n-1}$ , $c_{n-1}$. The algorithm continues until all Cauer $r$, $c$ are computed and is terminated at $p_0(s) = 0$.

It should be noted that the Cauer parameters $r_i$ , $c_i$ are determined unequivocally by the Foster-Cauer transformation indicating the physical meaning of the Cauer-circuit. On the other hand, the representation of the Foster impedances by the $R_i$ , $C_i$ is ambiguous, since every permutation of pairs of $R_i$, $C_i$ in (2) or Fig.2 leads to a mathematically identical expression.

In practice no more than 15 pairs of resistors and capacitors are necessary to represent a thermal impedance. In tests and applications one circuit representation was transformed into the other and then back again. In every case the original circuit parameters were recovered with perfect accuracy.

## IV. APPLICATION AND DISCUSSION

For packages of small interface size between power dissipating chip and cooling radiator it is not necessary to subdivide the interface in several pieces, as suggested at the end of section II, because the interface temperature is essentially homogeneous. We have performed thermal measurements with a TO-218AB package of about 2 cm² interface area mounted on a closed water cooled radiator. The package is shown in Fig. 6 and contains a single Siemens/Infineon 1200V/35A IGBT chip BUP 307.

The thermal impedance junction-ambient of the complete set-up, denoted by ZthJA(t), is the heating curve of the chip for applied constant dissipated chip power of unit strength (unit step-response). It is measured by monitoring the cooling down of the IGBT from a heated steady state after turn off of the heat generating IGBT. The temperature of the IGBT is observed by measuring the chip's on-state voltage for a small impinged constant current whose heat generation during cool down is negligible. ZthJA(t) is obtained from the cooling down curve by the superposition principle of the linear 3D-heat conduction equation [15]. This is valid under the assumption that material parameters (thermal conductivity, specific heat) in the set-up and boundary conditions are independent of temperature, but also applies for the typical material parameter variations in a range between 300°K and 400°K.

Fig. 7 shows the data points for ZthJA(t) inferred from the measured cooling down curve. The data points are arithmetic averages of the original digitalized measurement signals in small time intervals (one point for each interval). The curve fitted to the data was obtained with a nonlinear fit-routine working according to the Levenberg-Marquardt method [20] to fit the model (2) with 9 pairs $R_i$ , $C_i$ ($C_i = t_i /R_i$ ). Nearly the same fitting curve was obtained when fitting the original data with over 20,000 data points, with only small deviations at small times (below 1 ms) and temperatures where the measurement is not as accurate (Fig. 8).

The heat conduction equation leads in agreement with (3) to the convolution integral representation of the chip (junction) temperature for arbitrary chip-power dissipation profile $P(t)$:

$$T_J(t) - T_a = \int_0^t P(\tau)\, \dot{Z}thJA(t - \tau)\, d\tau \qquad (7)$$

($T_a$ = ambient temperature; $\dot{Z}thJA$ time derivative of ZthJA). Because of the direct derivation of (7) from the heat conduction equation, its validity is restricted to a certain class of boundary conditions and to homogeneous starting conditions $T(x, 0) = T_a$ as detailed in [15]. Also, (7) holds only, if the power density distribution in the chip is allowed to change its overall strength with time but not its spatial distribution.

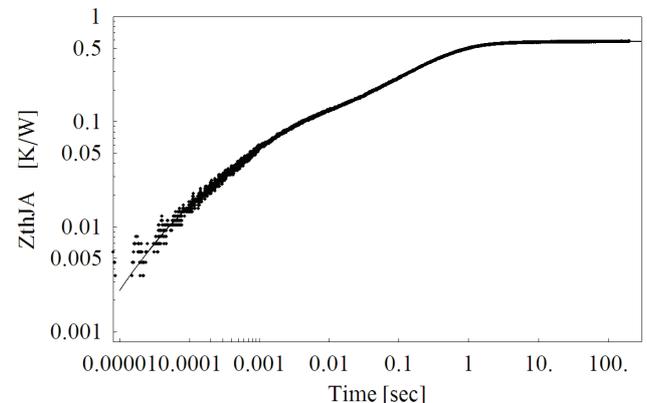

Fig 8: Same as in Fig. 7, but curve fit to original data with over 20,000 points.

Manufacturers data sheets for power devices usually contain only the thermal impedance of the package or module "junction to case" $ZthJC(t) = (T_J(t) - T_c(t))/P$, where the case temperature $T_c(t)$ depends on the selected location, e.g. in the middle of the interface between module-base-plate and cooling radiator. The thermal resistance of the interface (typically made up of a thermal grease) has to be included in the thermal impedance of the radiator. Sometimes in the literature $T_c(t)$ is measured simultaneously with $T_J(t)$ to obtain heating curves with constant heating. The $ZthJC(t)$ inferred from the difference $T_J(t) - T_c(t)$ is the difference of two monotonously increasing functions. However, this difference and the $ZthJC(t)$ thus defined are not necessarily monotonously increasing. Examples for nonmonotonous $ZthJC(t)$ in form of an "anomalous bump" have been observed in the literature [sof]. A meaningful definition of $ZthJC(t)$ for thermal characterisation of the package, which is independent of the choice of the cooling radiator or cooling conditions, can only be obtained for constant case temperature $T_c(t)$ = const. On the other hand, this condition is difficult to realise experimentally, because an ideal cooler would be required to keep the interface temperature constant in time. This could be achieved by an active thermoelectric cooler.

Our method proceeds in the following way and infers the unknown $ZthJC(t)$ for constant case temperature from the measured $ZthJA(t)$, when only the steady state case temperature (or the thermal resistance of the radiator) is known: The case temperature $T_c$ is measured in steady state before the cool down starts by a thermocouple pressed at the package bottom side through a hole in the radiator. The measured $ZthJA(t)$ is represented by the Foster-model (2) as described (Fig. 7). Then the Foster $R_i$, $C_i$ are transformed in Cauer $r_i$, $c_i$ with the algorithm of section III. The resulting Cauer ladder junction-ambient is cut into two pieces junction-case and case-ambient along the line shown in Fig. 9, where the sum of the resistors counted from the right hand (ambient) side becomes larger than $Rth_{radiator} = (T_c - T_a) / P$. The left part of the divided Cauer circuit is the $ZthJC(t)$, the Cauer circuit right from the cutting line forms the thermal impedance of the radiator $ZthCA(t)$.

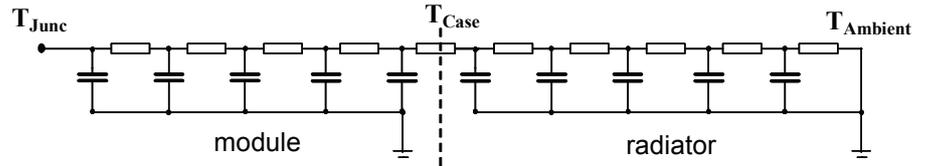

Fig. 9: Decomposition of Cauer-circuit for ZthJA into ZthJC and ZthCA along dashed line, where sum of resistors from ambient side = $Rth$-radiator. Also prescription for combining ZthJC and ZthCA to ZthJA.

It should be noted that the impedance for $ZthCA(t)$ consists of a Cauer ladder as in Fig. 3 together with the corresponding part of the thermal resistor divided by the cutting line attached to the left hand node. Due to this thermal series resistor the impedance $ZthCA(t)$ performs a step at $t = 0$ and rises at the beginning with infinite steepness contrary to $ZthJC(t)$ or $ZthJA(t)$ which rises linearly for $t < 10^{-4}$ sec in case of a volume heat source in the chip [11]. The heat source applied to the cooling radiator is no volume heat source in the radiator but a heat flux at its surface. In this case the temperature response to a unit step in the heat flux at the radiator surface behaves as $\sqrt{t}$ [22, 11]. Therefore the thermal series resistor in the model for $ZthCA(t)$ describes the physics more correctly, than the pure Cauer-circuit of Fig. 3. The transformed Foster-ZthCA has the same thermal series resistor as the Cauer-circuit.

The Cauer circuits thus obtained for $ZthJC(t)$ and $ZthCA(t)$ were transformed into Foster-circuits by use of the algorithm of section III. The Foster circuit elements - and also the Cauer elements - of $ZthJC(t)$ provide a suitable representation for the thermal characterisation in manufacturers data sheets. In order to obtain from the Foster values the complete impedance $ZthJA(t)$ of a combined system, the subsystem Foster-circuits have to be transformed in Cauer-circuits which are connected according to Fig. 9 at the interface node. After this the resulting Cauer-ladder junction-ambient is transformed in Foster-form for ease of mathematical presentation. Performing these steps we obtained the original Foster elements for $ZthJA(t)$ (Fig. 7) with perfect accuracy.

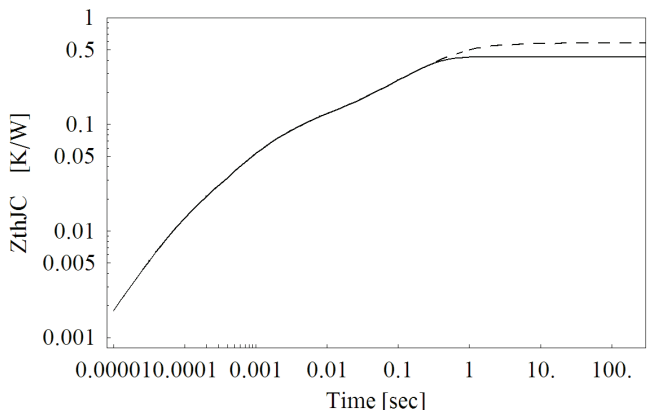

Fig. 10: ZthJC resulting from ZthJA (dashed line) of Fig. 7 according to the prescription of Fig. 9.

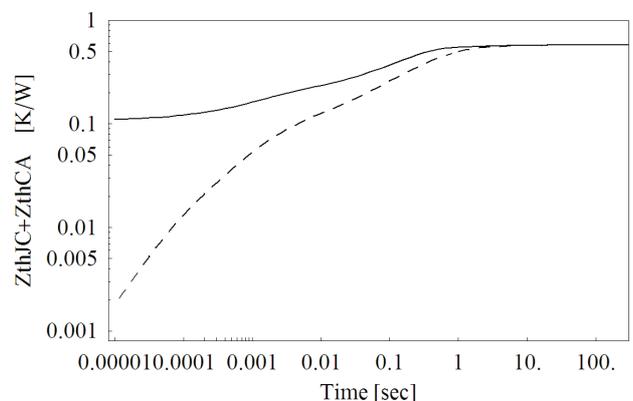

Fig. 11: Direct addition of the ZthJC, ZthCA constructed according to Fig. 9, leads to wrong prediction for ZthJA (dashed line).

The result of the construction of ZthJC(*t*) from ZthJA(*t*) by this prescription is displayed in Fig. 10 together with the original ZthJA(*t*). The difference ZthJA(*t*) - ZthJC(*t*) for large times is equal to the static thermal resistance of the radiator $Rth_{radiator}$.

The simple addition of the radiator impedance ZthCA(*t*) and of ZthJC(*t*) - which corresponds to a series connection of the Foster-circuits of the respective impedances - in order to obtain ZthJA(*t*) gives a grossly wrong result, as can be seen from Fig. 11. It is indispensable, first to transform the ZthJC, ZthCA in Cauer-ladders and to combine the Cauer-ladders to obtain the correct ZthJA.